# Effect of Dopamine in Enhancement of SNR of Cortico-Striatal-Thalamo-Cortical Loop Spiking

Hadi Barati[1], Ali Nayerifar[1], and Mehdi Fardmanesh[1]

[1]Department of Electrical Engineering, Sharif University of Technology, Tehran, Iran

Corresponding author:
Hadi Barati
Department of Electrical Engineering, Sharif University of Technology, Tehran, Iran
Email: baratihadi6@gmail.com , hadi.barati@ee.sharif.edu

**Abstract**
In this work, the effects of dopamine neurotransmitter within the Cortico-Striatal-Thalamo-Cortical (CSTC) loop have been investigated. Simulations confirmed dopamine facilitates movement via thalamic disinhibition. Analysis of its impact on the signal-to-noise ratio (SNR) revealed a complex, region-specific outcome: SNR increased in some regions (e.g., D2 Striatum: 3.41 dB to 6.25 dB), decreased in others (e.g., Thalamus VL: 6.24 dB to 3.93 dB), and remained stable elsewhere (e.g., M1: 3.16 dB to 3.13 dB). This heterogeneity stems from dopamine increasing the excitability of D1-receptor-expressing neurons, which amplifies channel conductance noise and reduces SNR in specific circuits. Thus, dopamine acts not as a uniform signal enhancer, but as a complex modulator that critically balances facilitation and noise within the CSTC loop.

**I. Introduction**

The dopamine (DA) has several roles in modulating the brain neural activity. In other words, it is crucial in regulating brain neural circuits to guide behavior, learning, motivation, motor control, etc. DA release as an error signal, differs between expected and actual rewards. This would be a powerful teaching mechanism for the brain [1]. DA is also necessary for motivation by assigning motivational values to cues, making them attractive. This vital for goal-directed responses [2]. DA profoundly influences prefrontal cortical (PFC) activity, which is essential for executive functions like working memory, cognitive flexibility, and attention [3]. However, the most classic role of DA is in regulating the movement commands sent from the motor cortex to the muscles. This regulation is performed through DA modulatory function in the striatum [4]. The increase in the signal-to-noise ratio (SNR) by dopamine is a fundamental and critically important mechanism for efficient neural computation, particularly in the prefrontal cortex (PFC) and striatum. It is not a minor side effect but a core principle of how dopamine optimizes brain function for goal-directed behavior. DA modulates PFC activity in an inverted U-shaped manner. Namely, the DA concentration should be in an optimum amount leading to an enhanced functionality of PFC with increased SNR. Too much or too low amount of DA release results in dysfunction of neuronal activity [5]. SNR increase has been reported in medial prefrontal cortex (mPFC) neurons in response to aversive stimuli. These neurons project to periaqueductal gray (PAG) located in midbrain [6]. by a novel SNR estimation method, it was found that single-neuron SNRs are very low (ranging from -29 dB to -3 dB) and that a neuron's intrinsic spiking history is often a more informative predictor of its activity than an external stimulus [7]. DA stabilizes neural signaling and its acute depletion in healthy individuals increases hemodynamic signal variability within motor and salience networks with decreased functional integration of these networks with the rest of the brain [8]. DA, by acting on $D_1$ receptors, helps stabilize neuronal activity, enhancing synaptic plasticity and reducing spontaneous firing through $D_2$ receptors leading to improved SNR towards appropriate behavior [9]. Two types of prefrontal cortex neurons can be differently affected by DA modulation: one type is quickly become underactive to maintain signal quality while another type is slowly overactive improving signal quality. Hence DA controls sensory influencing how the PFC processes sensory information [10]. Although some chemical substances such as nicotine can increase DA signaling through increasing frequency and duration DA neurons spiking, ventral striatum and dorsal striatum respond differently to this increase. The findings show that ventral striatum is significantly vulnerable to nicotine with a widen SNR [11]. DA neurons in ventral tegmental area have direct connections with neurons in the nucleus accumbens (ventral striatum), using glutamate as a fast excitatory signal. This signal can convey rapid incentive salience in addition to their slower dopaminergic actions [12].

In the present work, the modulatory effect of DA on different regions of the brain Cortico-Striatal-Thalamo-Cortical (CSTC) Loop such as motor cortices (PMC, M1), striatum (D1 and D2 receptors), etc. are investigated through simulations of the Hodgkin-Huxley model of the brain neural networks.

**II. Hodgkin-Huxley model**

The Hodgkin-Huxley (H-H) model is a mathematical model that calculated the action potential initiation and propagation in neurons. In this model, the equivalent electrical circuit of a neuron membrane is employed. In this electrical model, the ion channels and the dielectric effect of the membrane are included. The main equation stating that the total current flowing across the membrane is the sum of the capacitive and ionic currents, is presented as the following

$$C_m \frac{dV_m}{dt} + I_{leak} + I_{Na} + I_K + I_{syn} = I_{app} \tag{1}$$

$I_{app}$ is the applied electric or electrochemical current externally applied to the neuron (inside). $I_{leak}$, $I_{Na}$ ,and $I_K$ are ionic currents passing through the membrane. $I_{syn}$ is the synaptic current afferent to the neuron due to the synaptic connections of the presynaptic neurons. $C_m \frac{dV_m}{dt}$ is the polarizing current of the membrane that is due to charge accumulation or depletion near the membrane surface. $V_m$ is the transmembrane potential which equals to the membrane potential at inside the neuron minus its value at the outside the neuron. $C_m$ is the capacitance of the membrane. In the H-H formalism [13], the ionic currents can be presented a s the following equation

$$I = gm^a h^b (V_m - E) \tag{2}$$

The gating variable $m$ and $n$ corresponding to channel activation and inactivation, respectively and the exponents $a$ and $b$ indicates the number of the activation and inactivation gates. $g$ is the maximum electrical conductance of the ionic channel while $E$ is the Nernst's or reverse potential of the channel. In the same formalism, the differential equation for the dynamics of the gating variable, let's call it $x$, can be presented as follows

$$\frac{dx}{dt} = c(V_m) - \big(c(V_m) + d(V_m)\big)x \tag{3}$$

The parameters c and d have their own voltage-dependent equations [13]. For determining the total synaptic current to a neuron, first the synaptic activity $s$ of each presynaptic neuron is calculated through the equation [13]

$$\frac{ds}{dx} = (1-s)(1+\tanh(V_m)) - s \tag{4}$$

Then, the sum all synaptic currents are taken into account through the following relation

$$I_{syn} = g'(V_m - E') \sum_{i=1}^{N} s_i \tag{5}$$

$g'$ is the electrical conductance of the synaptic link. $E'$ is the reverse electrical potential of the synapse and $s_i$ is the synaptic activity of the presynaptic neurons.

Therefore, if the graph of the neuronal network is created and the presynaptic neurons of each neuron are determined, then by utilizing equations (1)-(5), the transmembrane potentials of the neurons can be calculated and be used for evaluating the network (graph) action potential evolution.

## III. Dopamine modulatory model

The dynamics of dopamine concentration in the extracellular space (synaptic cleft), $[DA]_e$, can be presented as the following equation [13]

$$\frac{d[DA]_e}{dt} = J_R - J_U - J_D \tag{6}$$

This equation takes into account three main processes in changing $[DA]_e$. Namely, $J_R$ is the flux of DA entering the synaptic cleft and is calcium dependent. $J_U$ is the flux of DA uptake back into the intracellular compartment of the presynaptic neuron. Finally, $J_D$ represents the DA removal from the cleft due to outward diffusion and degradation of DA molecules. The intracellular calcium ion concentration, i.e., $[Ca]_i$, at the synapse can be modeled by [14]

$$\frac{d[Ca]_i}{dt} = \alpha I_{Ca} - \beta([Ca]_i - [Ca]_{i,0}) \tag{7}$$

In this equation, $I_{Ca}$ is the high-threshold calcium current entering the presynaptic dopamine neuron and can be determined using the Hodgkin-Huxley formalism [15]

$$I_{Ca} = 3.275 \cdot \frac{1}{1+\exp[-(V_m+4.5)/5.2]} \cdot \frac{1}{1+\exp[-(V_m+74.8)/6.5]} (V_m - E_{Ca}) \tag{8}$$

In equation (8), $V_m$ is the transmembrane potential of the dopamine neuron. Hence $J_R$ can be calculated using the following equation [15]

$$J_R = \gamma \frac{[Ca]_i^4}{[Ca]_i^4 + K^4} \tag{9}$$

The dopamine uptake flux can be determined by [16]

$$J_U = \frac{\sigma[DA]_e}{k+[DA]_e} \tag{10}$$

The dopamine outward flux is also calculated through equation (10)

$$J_D = \epsilon[DA]_e \tag{11}$$

The values of the constants presented in equations (7)-(11) are brought to Table 1.

Table 1: Values of constants in equations (7)-(11)

| | |
|---|---|
| $E_{Ca}$ | 100 mV |
| $\alpha$ | 20728 nM/ms /(µA/cm^2µA/cm^2) |
| $\beta$ | 0.01 ms^−1 |
| $[Ca]_{i,0}$ | 100 nM |
| $\gamma$ | 244214.851 nM/ms |
| $K$ | 700000 nM |
| $\sigma$ | 6 nM/ms |
| $k$ | 30 nM |
| $\epsilon$ | 0.0083511 ms^−1 |

## IV. Simulation model

The motor circuit illustrated in the figure 1 represents a simplified schematic of CSTC loop, a critical network for the planning, initiation, and execution of voluntary movement. The circuit originates with excitatory glutamatergic projections from the primary motor cortex (M1) and premotor cortex (PMC) to the striatum, the main input nucleus of the basal ganglia. Within the striatum, information is processed through two primary pathways: the direct pathway, which facilitates movement via striatal projections expressing primarily D1 dopamine receptors that inhibit the internal segment of the globus pallidus (GPi) and the indirect pathway, which suppresses movement via striatal projections expressing primarily D2 receptors that inhibit the external globus pallidus (GPe), leading to disinhibition of the subthalamic nucleus (STN) and subsequent increased excitatory drive onto the GPi. The GPi serves as the primary output nucleus, providing tonic GABAergic inhibition to the motor thalamus (VA/VL). A reduction in this inhibitory output disinhibits the thalamus, allowing for thalamocortical excitation and the facilitation of movement.

The dopaminergic pathway originating from the substantia nigra pars compacta (SNc) plays a pivotal modulatory role in this circuit, essential for normal motor function. Dopamine released in the striatum differentially modulates the direct and indirect pathways by exciting D1 receptor-expressing neurons of the direct pathway and inhibiting D2 receptor-expressing neurons of the indirect pathway. This dual action synergistically promotes motor activity by reducing the inhibitory output from the GPi. The degeneration of dopaminergic neurons in the SNc, as seen in Parkinson's disease, leads to a profound imbalance in these pathways. The loss of dopamine results in overactivity of the indirect pathway and underactivity of the direct pathway, culminating in excessive inhibition of the thalamus and suppressed motor cortex activity, which manifests clinically as bradykinesia, rigidity, and tremor. For simplicity, in the present work, SNc is modeled by a single neuron and H-H equations are used for calculating $V_m$ used in equation (8). The SNc neuron is excited by an external applied current.

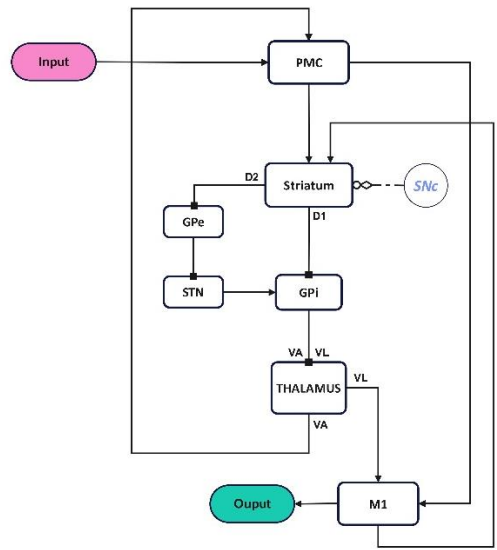

Figure 1. Cortico-basal ganglia-thalamo-cortical (CSTC) loop

Based on the illustrated graph in Figure 2, the CSTC loop is represented as a complex network, generated using the Watts-Strogatz small-world algorithm. This computational approach effectively captures the architectural principle of high local clustering with short path lengths, a hallmark of neural systems that balances specialized processing with efficient global integration. The synaptic connectivity within and between these regions is specifically defined based on established neurobiological literature [17-20], ensuring that the fundamental excitatory and inhibitory pathways of the CSTC loop are accurately reconstructed. The directions of the synaptic connections for each region are also depicted in Figure 2.

The dynamical behavior of each neural population within these regions is simulated using the biologically realistic H-H formalism, as specified by equations (1)-(5). This framework models the generation of action potentials by tracking the flow of sodium, potassium, and other ions across neuronal membranes, providing a high-fidelity representation of neuronal excitability and spiking patterns. Crucially, each region is populated by a mixture of principal excitatory or inhibitory neurons and local inhibitory interneurons, with the specific cell types assigned according to published anatomical and physiological studies. This heterogeneity is essential for generating the rich repertoire of oscillatory activities and state transitions observed in the basal ganglia-thalamocortical system. The synaptic links between neurons, both within a node and across connected nodes, are established based on these defined neuron types, creating a network where excitation and inhibition are carefully balanced to support stable yet flexible dynamics. For the simulations of the present work, the neuronal numbers of CSTC components are set as M1=10, PMC=100, D1 striatum=100, D2 striatum=100, GPe=50, GPi=50, STN=50, VA=100, and VL=100. The values of the constants appearing in H-H equations are set as reported in [18-21]. However, the synaptic weights in the synaptic links illustrated in Figure 2, are defined in such a way that the synaptic activities of the lined regions become appropriately correlated. Namely, the effective connectivity leads to the functional connectivity for the linked areas.

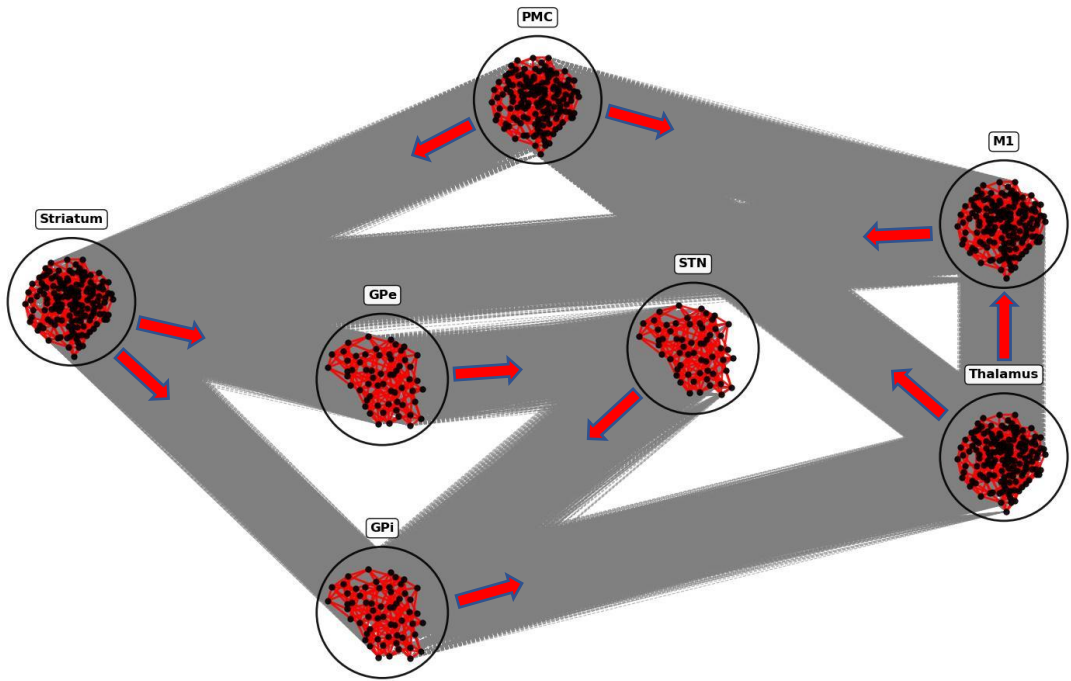

Figure 2. Small world graph of CSTC loop

A critical component of the model is the incorporation of dopamine modulation within the Striatum, governed by equations (6)-(11). Dopamine is known to critically regulate the gain of cortical signals through the striatal medium spiny neurons (MSNs), differentially modulating the direct and indirect pathways. In other words, the dopamine increases the Na channel conductivity of D1-receptor MSN and increase the K channel conductivity of D2-receptor MSN. In this model, dopamine influences synaptic efficacy, neuronal excitability, and plasticity, thereby acting as a key control parameter that can shift the operational mode of the entire CSTC loop. By integrating this neuromodulation into a small-world network of H-H neurons, the model offers a powerful computational platform to investigate the circuit-level mechanisms underlying normal cognitive functions, such as action selection and learning, as well as the pathophysiological states that arise from dopaminergic dysregulation, such as those observed in Parkinson's disease. For simplicity, here, the dopamine neuronal population is just modeled by a single neuron and its spiking activity is applied equally to all neurons of its targets, viz., striatum D1 and D2 neurons.

## V. Results and discussion

In Figure 3, the spiking activities of the CSTC regions with no noise and DA inclusions are illustrated. As expected, due to lack of DA, the D1 neurons are not enough activated to inhibit GPi. In addition, the D2 neurons are not inhibited and thus the STN neurons remain active and they spike to excite the GPi neuron that leads to more activity of the GPi neurons. A high excited GPi inhibits the thalamus which results in movement stops, i.e. M1 and PMC neurons are not appropriately spiking. It should be noted that for all regions, a minimum synaptic activity, known as intrinsic or innate activity, is established by applying an external current to the H-H model, i.e., $I_{app}$ in Equation (1). In addition, the regions are more excited if they are active in real situations. For example, the M1 and PMC regions are excited when the movement is intended.

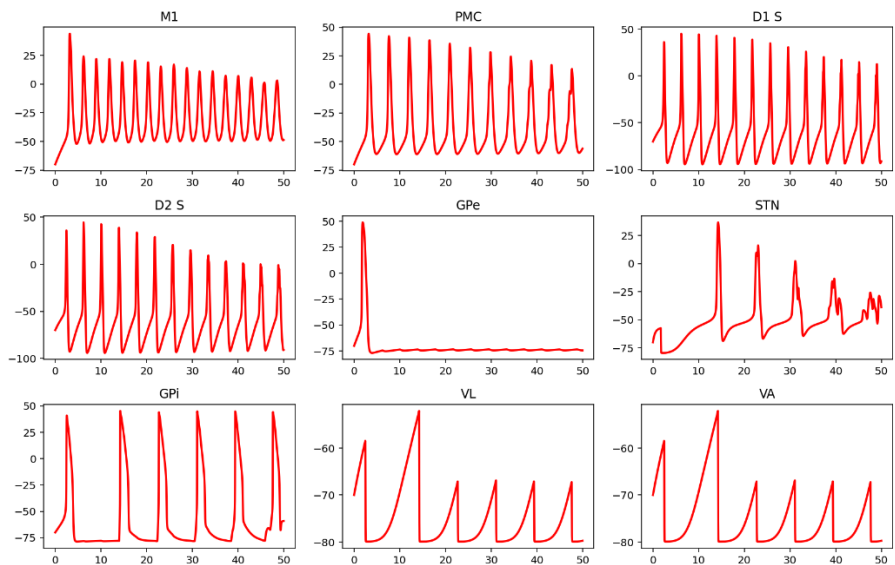

Figure 3. Spiking activities of CSTC regions when noise and DA are not applied to neurons. The vertical axis is transmembrane potential in mV and the horizontal axis is tie in ms.

Now, the effects of DA modulation on the CSTC loop components spiking dynamics without any noise introduced to the system are investigated. The results are shown in Figure 4. As expectedly can be seen in this figure, DA plays a crucial role in facilitating voluntary movement through a key disinhibition mechanism within the CSTC circuit. DA excites the "direct pathway" by binding to D1 and D2 receptors, which facilitates the inhibition of the basal ganglia's output nuclei (GPi). These output nuclei normally exert a tonic, inhibitory "brake" on the thalamus. By suppressing this inhibitory output, dopamine effectively disinhibits the thalamus. This release of the thalamic brake allows the thalamus to freely excite the motor cortices M1 and PMC, thereby initiating and sustaining the neural commands necessary for smooth, voluntary movement.

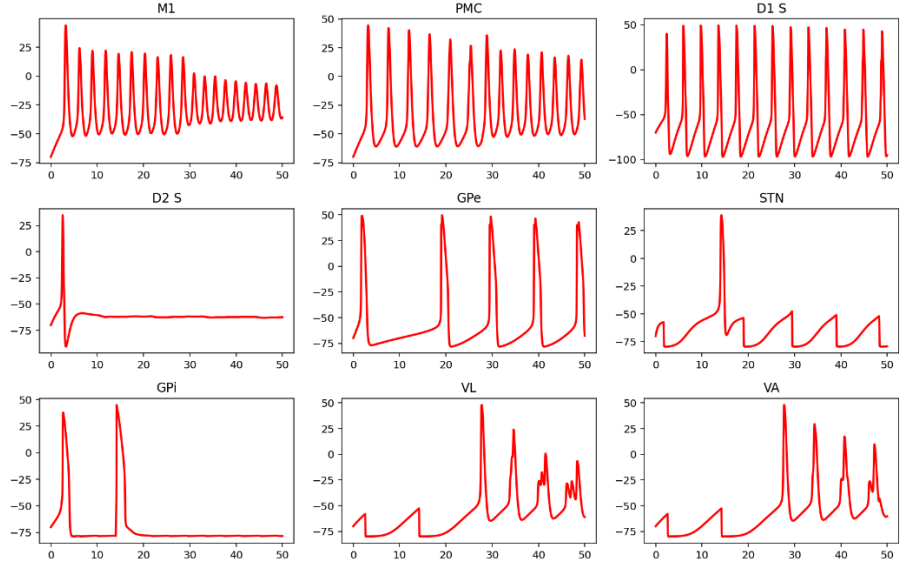

Figure 4. Spiking activities of CSTC regions when noise is not applied but DA modulation is included. The vertical axis is transmembrane potential in mV and the horizontal axis is tie in ms.

The enhancing influence of DA on SNR of loop spiking is studied by imposing noise to the system. The noise inclusion is performed as described in the following [21]. The noise directly represents fluctuations in ionic conductances due to stochastic channel openings and closings. This implementation follows the conductance noise approach described in the [21] applied to H-H equation:

$$C_m \frac{dV_m}{dt} + I_{leak} + I_{Na,noisy} + I_{K,noisy} + I_{syn} = I_{app} \qquad (12)$$

In equation (12), $I_{Na,noisy} = g_{Na}(m^3 h + \xi_{Na})(V_m - E_{Na})$ and $I_{K,noisy} = g_K(m^4 + \xi_K)(V_m - E_K)$. $\xi_{Na}$ and $\xi_K$ are noise terms which are assumed as white noise with Gaussian distribution function [17]. The simulation results are brought to table 1. Based on the provided table, the analysis of the SNR values across different brain regions, both without and with DA, reveals a varied and region-specific impact. The D1 Striatum shows a decrease in SNR from 3.44 dB without DA to 2.52 dB with DA. In contrast, the D2 Striatum demonstrates a substantial improvement, with its SNR increasing from 3.41 dB to 6.25 dB. M1 exhibits minimal change, moving from 3.16 dB to 3.13 dB, while PMC shows a decrease from 3.16 dB to 2.56 dB. The GPi shows a positive change, with its SNR increasing from 3.87 dB to 4.9 dB. Conversely, GPe shows a notable decrease from a relatively high value of 5.94 dB without DA down to 4 dB with DA. The STN experiences a slight decrease from 4.92 dB to 4.68 dB. The most pronounced negative changes are observed in the thalamic regions. The VL Thalamus shows a significant reduction in SNR, falling from 6.24 dB without DA to 3.93 dB with DA. Similarly, the VA Thalamus decreases from 6.23 dB to 4.03 dB. Overall, the introduction of Dopamine has a dichotomous effect, substantially enhancing the signal quality in the D2 Striatum and moderately in the GPi, while being detrimental in other regions, particularly the thalamic areas and the D1 Striatum. Therefore, in due to more excitability added by DA in D1 neurons, since the noises are channel conductance noises, the noise amplitude increases leading to low SNR.

Table1. Influence of DA modulation on SNR

| Brain region | SNR (without DA) (dB) | SNR (with DA) (dB) |
|---|---|---|
| D1 Striatum | 3.44 | 2.52 |
| D2 Striatum | 3.41 | 6.25 |
| M1 | 3.16 | 3.13 |
| PMC | 3.16 | 2.56 |
| GPi | 3.87 | 4.9 |
| GPe | 5.94 | 4 |
| STN | 4.92 | 4.68 |
| Thalamus VL | 6.24 | 3.93 |
| Thalamus VA | 6.23 | 4.03 |

**VI. Conclusion**

In conclusion, this work has systematically investigated the multifaceted effects of dopamine within the CSTC loop. The initial phase of the study, utilizing H-H model simulations, confirmed dopamine's modulatory role in facilitating movement by implementing a disinhibition mechanism of the thalamus. Subsequently, the investigation into dopamine's impact on SNR across CSTC regions revealed a complex, region-specific outcome: SNR values increased in some areas, decreased in others, and remained largely unchanged elsewhere. This heterogeneity can be attributed to the increased excitability conferred by dopamine on D1-receptor-expressing neurons. Since the predominant noise source is channel conductance noise, the heightened neuronal activity consequently amplifies noise amplitude, leading to a reduction in SNR in specific circuits. Therefore, this research elucidates that dopamine's role is not uniformly beneficial for signal fidelity but is instead a nuanced modulator that balances facilitation and noise, critically shaping information processing within the CSTC loop.